\documentclass[singlecolumn,superscriptaddress,aps,prl,preprint]{revtex4-1} 

\usepackage{graphicx}
\usepackage{color}
\usepackage[colorlinks=true,
            linkcolor=blue,
            urlcolor=blue,
            citecolor=blue,
            hyperfootnotes=true]{hyperref}

\begin{document}
\title{Saddle-like topological surface states on the TT'X family of compounds (T, T' = Transition metal, X= Si, Ge)}

\author{Bahadur Singh}
\affiliation{Centre for Advanced 2D Materials and Graphene Research Centre, National University of Singapore, Singapore 117546}
\affiliation{Department of Physics, National University of Singapore, Singapore 117542}

\author{Xiaoting Zhou}
\affiliation{Centre for Advanced 2D Materials and Graphene Research Centre, National University of Singapore, Singapore 117546}
\affiliation{Department of Physics, National University of Singapore, Singapore 117542}

\author{Hsin Lin}
\email{nilnish@gmail.com}
\affiliation{Centre for Advanced 2D Materials and Graphene Research Centre, National University of Singapore, Singapore 117546}
\affiliation{Department of Physics, National University of Singapore, Singapore 117542}

\author{Arun Bansil}
\affiliation{Department of Physics, Northeastern University, Boston, Massachusetts 02115, USA}

\begin{abstract}
Topological nodal-line semimetals are exotic conductors that host symmetry-protected conducting nodal-lines in their bulk electronic spectrum and nontrivial drumhead states on the surface. Based on first-principles calculations and an effective model analysis, we identify the presence of topological nodal-line semimetal states in the TT'X family of compounds (T, T' = transition metal, X= Si, or Ge) in the absence of spin-orbit coupling (SOC). Taking ZrPtGe as an exemplar system, we show that this material harbors a single nodal line on the $k_y=0$ plane, which is protected by the $M_y$ mirror plane symmetry. Surface electronic structure calculations further reveal the existence of a drumhead surface state nested inside the nodal line projection on the (010) surface with a saddle-like energy dispersion. When the SOC is included, the nodal line gaps out and the system transitions to a strong topological insulator state with $Z_2=(1;000)$. The topological surface state evolves from the drumhead surface state via the sharing of its saddle-like energy dispersion within the bulk energy gap. These features differ remarkably from those of the currently known topological surface states in topological insulators such as Bi$_2$Se$_3$ with Dirac-cone-like energy dispersions.  
\end{abstract}

\maketitle
Recent discovery of non-trivial band structures in semimetals 
expands topological classification beyond the insulators \cite{TopologicalBTh,Report_balents2011,Report_Jia2016,Review_Ashvin2013,Review_hasan2015,Review_nlFang2016,weyl_ashwin2011,weyl_singh2012,nodal_th2011,nodal_thT2014,nodal_thChiu2014,nodal_Fu2015, dirac_Liu12014,dirac_Liu22014,weyl_SMH2015,weyl_Weng2015,weyl_Xu2015,weyl_Lv2015}. Unlike topological insulators (TIs), which have well-defined band gaps in their bulk electronic spectra with interesting topological states only on their surfaces, topological semimetals (TSMs) feature unusual crystal-symmetry-protected states both in the bulk as well on the surface. In particular, the bulk Fermi surfaces of TSMs enclose non-trivial band-touching points that bring quantized numbers for integral of the Berry flux over any closed surface enclosing these points, and provide the relevant topological invariants. Depending on band degeneracy and the momentum-space distribution of the band-touching points, three different types of TSMs, namely, Weyl semimetal (WSM)\cite{weyl_ashwin2011,weyl_singh2012,weyl_SMH2015,weyl_Lv2015,weyl_Weng2015,weyl_Xu2015}, Dirac semimetal (DSM)\cite{dirac_Liu12014,dirac_Liu22014}, and nodal-line semimetal (NLSM)\cite{Review_nlFang2016,nodal_thT2014,nodal_thChiu2014,nodal_Fu2015} have been proposed. WSMs and DSMs exhibit two- and four-fold band-touching points with low energy Weyl and Dirac fermion excitations, respectively, with zero dimensional (0D) Fermi surfaces the 3D bulk Brillouin zone (BZ). In sharp contrast to WSMs and DSMs, NLSMs support extended band-touching points along a line with 1D Fermi surfaces in the bulk BZ. 
WSMs have been found in noncentrosymmetric TaAs\cite{weyl_Xu2015,weyl_Lv2015}, Mo$_x$W$_{1-x}$Te$_2$\cite{weyl_mote2,weyl_moxwte2}, LaAlGe\cite{weyl_LaAlGe} families, while DSMs have been realized in Na$_3$Bi\cite{dirac_Liu12014} and Cd$_2$As$_3$\cite{dirac_Liu22014}.

Among the other members of TSMs, NLSMs 
offer many unique properties that are distinct from WSMs and DSMs\cite{nodal_th2011,nodal_thT2014,nodal_thChiu2014,nodal_Fu2015,weyl_floquet,nodal_flatSC2011,nodal_inter2016}. For example, NLSMs have higher density of states at the Fermi level than DSMs and WSMs, and therefore, provide an ideal platform to study interaction-induced instabilities. They also feature topological surface states, known as ``Drumhead surface states (DSSs)'', which could be interesting for achieving superconductivity and correlation physics\cite{nodal_flatSC2011,nodal_inter2016}. The existence of NLSM state has been proposed theoretically in many materials including TaAs\cite{weyl_SMH2015,weyl_Weng2015}, antiperovskite Cu$_3$Pd(Zn)N\cite{nodal_thKim2015,nodal_thRui2015}, photonic crystals\cite{nodal_gyroid2013}, TlTaSe2\cite{nodal_TlTaSe2}, 3D carbon networks\cite{nodal_Wang2016,nodal_Mullen2016}, and other materials\cite{nodal_thYA2016,nodal_thChan2016,nodal_BaSn2}. The experimental evidence of NLSMs is however reported recently for PbTaSe$_2$\cite{nodal_expBian2016}, ZrSi(S,Te)\cite{nodal_expZrSiS1,nodal_expZrSiS2}, and PtSn$_4$\cite{nodal_PtSn4}. A focus of discussion has been the issue of the stability of nodal lines in the absence of spin-orbit coupling (SOC) effects\cite{nodal_thChiu2014,nodal_Fu2015,nodal_thKim2015,nodal_thRui2015,nodal_thChan2016,nodal_thYA2016}. Turning on the SOC either splits the nodal line into nodal points (Dirac points or Weyl points), depending upon the crystalline symmetries present,\cite{weyl_SMH2015,weyl_Weng2015,nodal_thKim2015,nodal_thRui2015} or fully gaps it due to the hybridization of bands with same symmetry eigenvalues\cite{nodal_thYA2016,nodal_thChan2016,nodal_BaSn2,nodal_expZrSiS1,nodal_expZrSiS2}. Regardless, saddle-like topological surface states have not been found in these materials.

In this work, based on systematic band structure calculations and model Hamiltonian analysis, we identify topological nodal-line fermions state in the large family of earth-abundant silicides or germanides, TT'X (T, T' = Transition metal, X= Si, or Ge), when the SOC is ignored. Taking ZrPtGe as an explicit example, we show that a single nodal line with large energy-momentum dispersion lies on the $k_y$=0 bulk plane and a saddle-like DSS nested inside the nodal line projection on (010) surface. Inclusion of the SOC gaps the nodal line, realizing a $Z_2$ nontrivial topological state with $Z_2=(1;000)$. The topological surface state evolves from the DSS with similar saddle-like energy dispersion. This is very unique since the known TIs such as Bi$_2$Se$_3$ have Dirac-cone-like surface states\cite{bi2se3_expxia,Hexawarp_Fu}. It is well known that saddle-points in the band structure give rise to interesting saddle-point Van Hove singularities (VHSs) where the density of states diverges in the 2D space \cite{vhs1953}. Once the VHS lies close to the Fermi level, the instabilities among lattice, charge, and spin degrees of freedom as well superconducting transition temperature, ferromagnetism and/or antiferromagnetism are substantially enhanced even in the weak interaction limit\cite{vhs_SC1965,vhs_SCinst2001,vhs_FM,vhs_FMP}. Our proposal of TNSM with saddle-like energy dispersed surface states in the large family of silicides and germanides therefore provides an exciting materials platform to explore these exotic properties in the presence of nontrivial band topology.

We perform electronic structure calculations within the density functional theory (DFT)\cite{kohan_dft} framework with the projector augmented wave (PAW) method\cite{vasp,paw} and generalized gradient approximation (GGA) \cite{pbe} energy functional, using the Vienna Ab Initio Simulation Package (VASP) code\cite{vasp}. SOC is included self-consistently to consider the relativistic effects. We use experimental lattice parameters with an energy cut-off of 350 eV for the plane-wave basis set and a tolerance of $1.0\times 10^{-8}$ eV for electronic energy minimization. The surface energy dispersions are calculated within the tight binding scheme based on the maximally localized Wannier functions (MLWFs)\cite{wannier90}, using the Wannier-tools software package\cite{WT_code,WT_Sancho1985}. 

\begin{figure}[t] 
\includegraphics[width=1.0\textwidth]{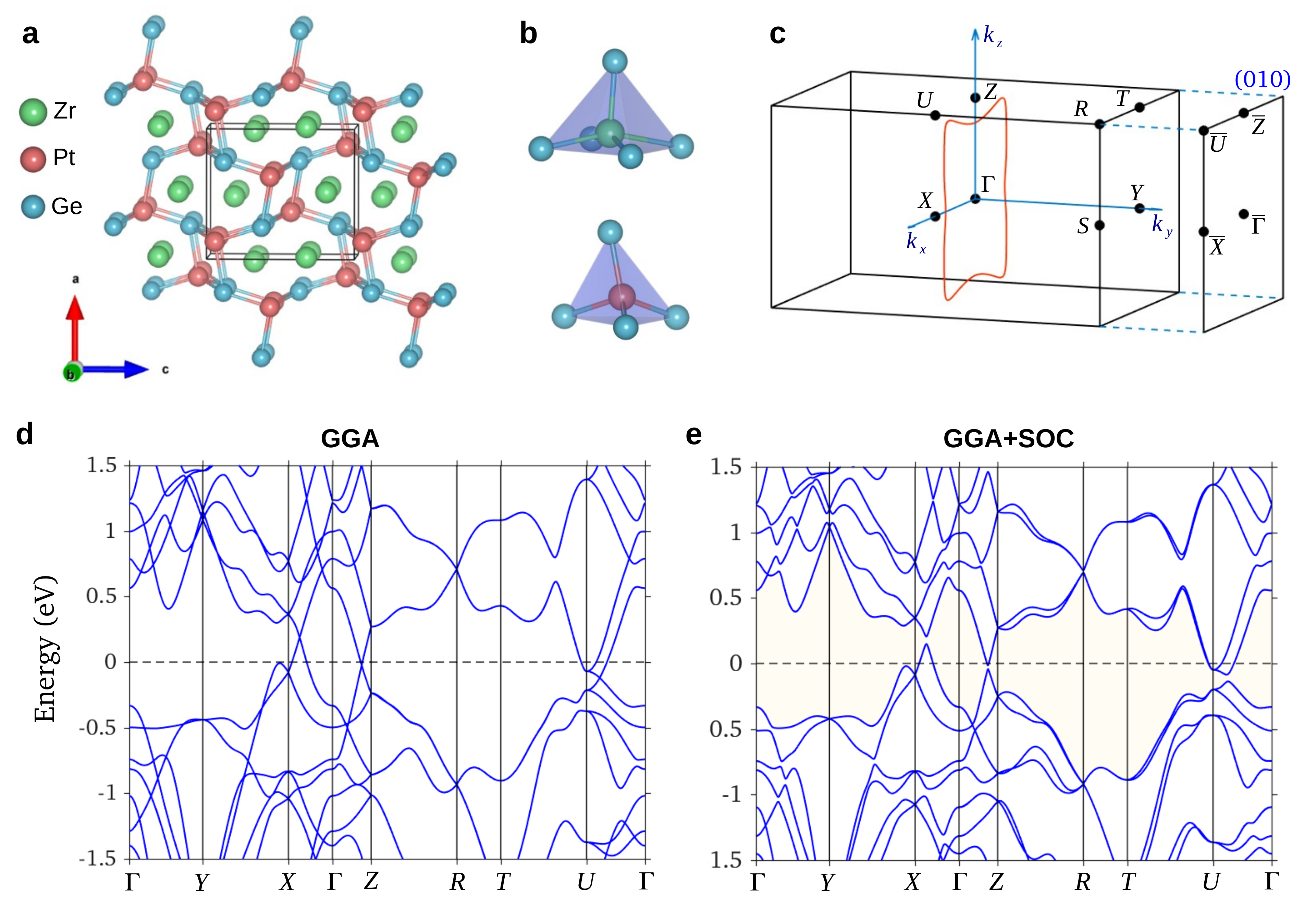} 
\caption {Crystal lattice and bulk bands of ZrPtGe. (a) Crystal structure of orthorhombic ZrPtGe with $Pnma$ (No. 62) symmetry.  Green, red, and blue balls denote Zr, Pt, and Ge atoms, respectively. (b) Local coordination structure of Zr (top panel) and Pt (bottom panel) atom in the unit cell. (c) Bulk Brillouin zone with its projection on the (010) surface. The relevant high symmetry points are marked with black dots. The calculated momentum distribution of nodal line is shown with orange color on the $k_y$=0 plane. (d) First-principles bulk band structure without the inclusion of spin-orbit coupling (SOC). Band crossings are evident along the $\Gamma-X$, $\Gamma-Z$, and $\Gamma-U$ directions near the Fermi level. (e) Bulk band structure with the inclusion of SOC. A small band gap opens at the band-crossing points. The shaded region highlights the continuous gap below which the topological invariants are calculated.}
\label{fig:lattice}
\end{figure}

We start with the discussion of crystal structure of TT'X (T, T' = Transition metal, X= Si, Ge) materials. All these compounds crystallize in a orthorhombic Bravais lattice with the non-symmorphic space group $D^{16}_{2h}$ ($Pnma$, No. 62) \cite{ZrPtGe_CS1,ZrPtGe_CS2}. Each compound contains an early transition metal as one component, a late transition metal as another component along with either the silicon or the germanium atom. The crystal structure of ZrPtGe is illustrated in Fig. \ref{fig:lattice}(a) as an example. In this structure, Pt and Ge atoms form strongly corrugated Pt$_3$Ge$_3$ hexagonal networks and Zr atoms fill the cavities left in these networks. Due to the strong puckering between different atomic layers, Pt forms a distorted tetrahedral configuration with Ge whereas Zr is coordinated to five Ge atoms as noted in Fig. \ref{fig:lattice}(b). The associated crystal symmetries include an inversion center $i$, two-fold screw rotation axes $\{C_{2x}|\frac{1}{2} \frac{1}{2}\frac{1}{2}\}$,  $\{C_{2y}|0\frac{1}{2}0\}$, and $\{C_{2z}|\frac{1}{2} 0 \frac{1}{2}\}$, and three glide-mirror planes $\{M_{x}|\frac{1}{2}\frac{1}{2}\frac{1}{2}\}$, $\{M_{y}|0\frac{1}{2}0\}$, and $\{M_{z}|\frac{1}{2} 0 \frac{1}{2}\}$. Additionally, all system respect the time reversal symmetry $\Theta$. Among these symmetries, $\{M_{y}|0\frac{1}{2}0\}$ represents a simple reflection around $y=0$ plane since its fractional translation is removable with a different choice of the origin. It plays the key role in protecting the nodal lines in present systems in the absence of SOC.  The first orthorhombic bulk and (010) surface BZs with the relevant high symmetry points are shown in Fig. \ref{fig:lattice}(c).

In Fig. \ref{fig:lattice}(d) we present the electronic structure of ZrPtGe without considering SOC. The valence and conduction bands cross along the high-symmetry lines $\Gamma-X$, $\Gamma-Z$, and $\Gamma-U$ that are tied to $k_y=0$ plane of the bulk BZ. A full BZ exploration shows that these band crossings persist along a closed path, realizing a single nodal line on the $k_y=0$ plane inside the bulk BZ as shown in Fig. \ref{fig:lattice}(c). This indicates clearly that ZrPtGe is a NLSM without SOC. As we consider SOC, the nodal line evaporates with the opening of a gap at the band crossings points as illustrated in Fig. \ref{fig:lattice}(e). It should be noted that owing to the coexistence of  $i$ and $\Theta$, each band still remains doubly degenerate in the presence of SOC at each $k$-point. Furthermore, additional non-symmorphic crystalline symmetries in this system lead to four-fold band crossings at the BZ boundary planes ($k_i=\pi$ planes) above or below the Fermi level. These band crossings are protected against gap opening and may realize high-symmetry Dirac cones at $X$, $Y$, and $Z$ points. The energy-dispersion can be tuned by changing the transition metal elements in the system as shown in Supplementary Materials.

\begin{figure}[t] 
\includegraphics[width=1.0\textwidth]{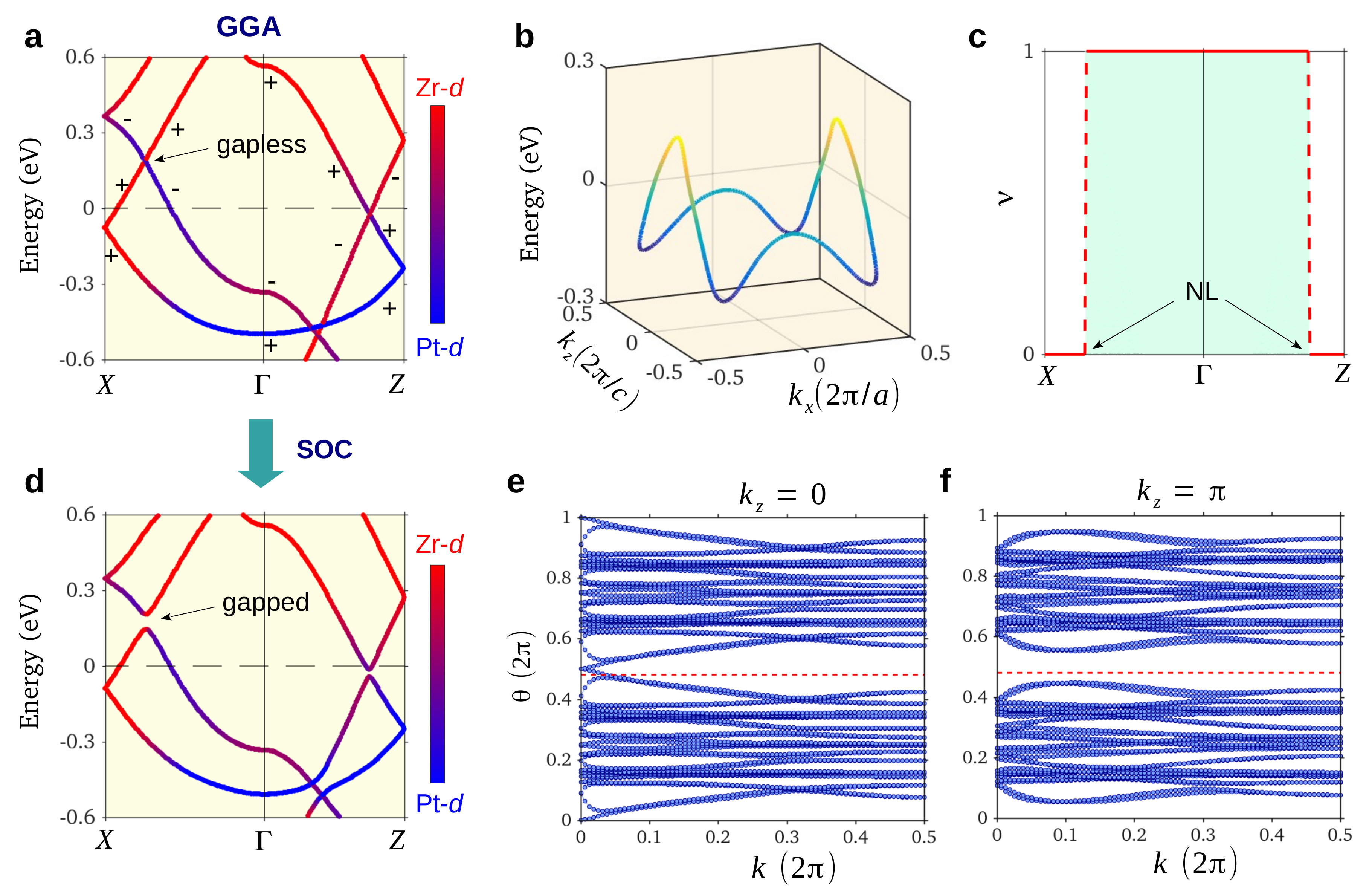} 
\caption{Nodal fermion and $Z_2$ topological insulating state in ZrPtGe. (a) Orbital compositions and mirror eigenvalues of the bulk bands without SOC. (b) Energy-dependent nodal line configuration in $E-k_x-k_z$ space on the $k_y$=0 plane. (c) Variation of non-trivial topological invariant along high-symmetry lines in the bulk Brillouin zone. The topological invariant is calculated using $M_y$ mirror eigenvalues of the occupied bands on mirror symmetric $k_y=0$ and $k_y=\pi$ planes without SOC. (d) Orbital compositions of the bulk bands with SOC. A small band gap opens at the band crossings as pointed by arrow. (e)-(f) Evolution of Wannier charge centers (WCCs) for $k_z=0$ and $k_z=\pi$ planes. WCCs evolution lines cross arbitrary reference line (red broken line) an odd number of times on $k_z$ = 0 plane, resulting in $Z_2$ = 1.
 }
\label{fig:nodalstr}
\end{figure}

In order to characterize the nature and the topological protection of nodal line, we systematically examine the band crossings in Fig. \ref{fig:nodalstr}. We know that nodal line resides on $k_y=0$ plane which is a $M_y:  (x,y,z)\rightarrow(x,-y,z)$ mirror invariant plane. Each band on this plane can have a well defined $M_y$ mirror eigenvalue. If a band crossing happens between two bands of different eigenvalues, it can remain gapless. Figure  \ref{fig:nodalstr}(a) shows the bands along two principal directions on the $k_y=0$ plane and corresponding mirror eigenvalues in the absence of SOC. Note that we have obtained $M_y$ eigenvalues from the first-principles Bloch wavefunctions and since the Hamiltonian remains spin-rotation invariant without SOC, it has eigenvalues $+1$ or $-1$. It can be seen that the lowest conduction band and the highest valence band have different mirror eigenvalues and thus their crossing points are topologically protected against gap opening. Further analysis of orbital character shows that bands around Fermi level are dominated by Pt $d$ and Zr $d$ states with a clear signature of band inversion at the $\Gamma$-point [see Fig. \ref{fig:nodalstr}(a)]. This implies a nontrivial band topology in the system even without SOC. We explicitly calculated the topological invariant for system using the mirror eigenvalues analysis\cite{nodal_thYA2016,nodal_thChan2016}. The computed topological invariant $\nu$ along the high-symmetry lines is presented in Fig. \ref{fig:nodalstr}(c). It takes nontrivial value only inside the nodal line and thus, signals the existence of odd number of DSSs inside the nodal line projection over the surface\cite{nodal_thYA2016,nodal_thChan2016}. In Fig. \ref{fig:nodalstr}(b) we present the nodal line structure in the $E-k_x-k_z$ space. The nodal line has a big energy dispersion around the Fermi level with corresponding energy values of 0.189 eV, -0.023 eV, and -0.119 eV along $k_x$, $k_z$, and plane diagonal ($k_x-k_z$) directions, respectively.

Figure \ref{fig:nodalstr}(d) illustrates the energy bands with SOC. In general SOC can drive NLSMs into DSM, WSM, or a full gap insulator. However considering the lower crystalline symmetries of ZrPtGe, SOC has opened the full gap at the crossing points, 
making the conduction band and the valence band to separate at each k-point. As we know that nodal line winds around a single time reversal invariant point ($\Gamma$-point) with nontrivial band topology, it is possible that SOC drives the system into a nontrivial insulating state. We have checked this by explicitly calculating the Wannier charge centers flow in the time reversal invariant planes $k_i=0,\pi$; results of two representative planes are shown in Figs. \ref{fig:nodalstr}(e)-\ref{fig:nodalstr}(f). These results show that ZrPtGe becomes a strong TI with $Z_2$=(1;000).

\begin{figure}[t] 
\includegraphics[width=1.0\textwidth]{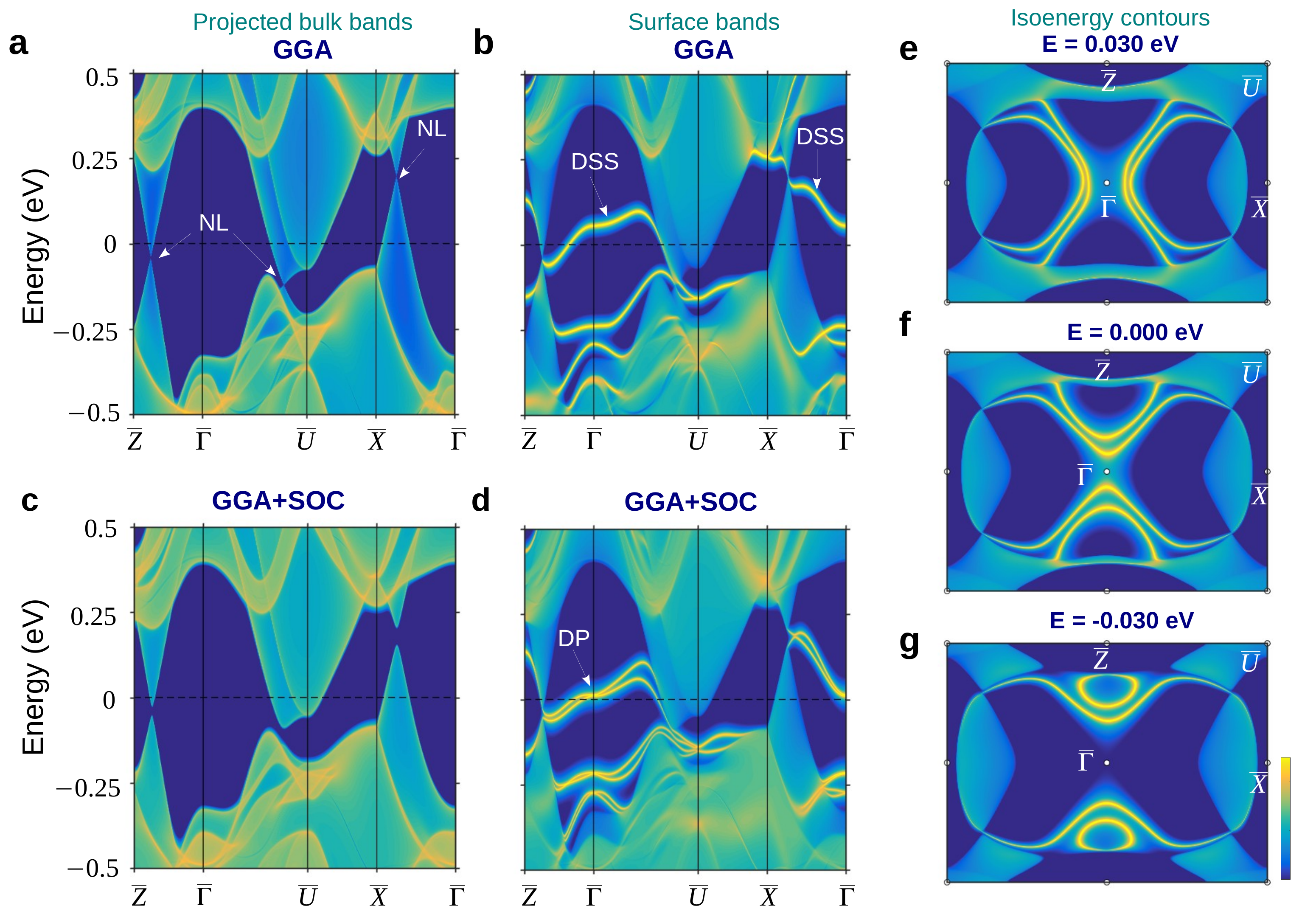} 
\caption{Topological electronic states on (010) surface of ZrPtGe. (a) Projected bulk bands onto (010) surface without SOC. The nodal line (NL) crossings are pointed with arrows. (b) Band structure of (010) surface without SOC. The non-trivial drumhead surface states (DSSs) are marked with arrows. (c)-(d) Same as (a)-(b) but with the inclusion of SOC. The non-trivial surface states evolve from the DSSs with different band curvature along different high-symmetry directions. The Dirac point (DP) is identified with white arrow. Isoenergy band contours at (f) E = 0.030 eV, (g) E = 0.000 eV, and (h) E = -0.030 meV with SOC. }
\label{fig:surface010}
\end{figure}

One important signature of topological NLSM is the existence of DSS either inside or outside the nodal line projection. Figure \ref{fig:surface010} shows the states for semi-infinite (010) surface of ZrPtGe which is parallel to bulk mirror plane $M_y$. The bulk bands projected onto (010) surface without SOC are shown Fig. \ref{fig:surface010}(a) where the nodal-line crossings are clearly seen. The DSSs nested inside the nodal line are visible in Fig. \ref{fig:surface010}(b) which is consistent with the calculated non-trivial invariant inside the nodal line.  Unlike the nearly flat DSSs reported in earlier works\cite{nodal_thKim2015,nodal_thRui2015,nodal_thYA2016,nodal_thChan2016,nodal_BaSn2}, the states of ZrPtGe are more dispersive, and interestingly, they have opposite band curvatures along $\overline{\Gamma}-\overline{X}$ and $\overline{\Gamma}-\overline{Z}$ directions, realizing an unique saddle-like energy-momentum dispersion relation. Such a 2D saddle-like bands are proposed as a route to achieve many fascinating properties\cite{nodal_flatSC2011,vhs_FM,vhs_SC1965,vhs_SCinst2001,vhs_FMP}.

Figures \ref{fig:surface010}(c) and \ref{fig:surface010}(d) show (010)-projected bulk bands and surface bands, respectively, with the inclusion of SOC. The nodal line is now gapped and DSS splits away from the time reversal invariant $\overline{\Gamma}$-point, deforming into topological Dirac-cone state. Since SOC is much smaller than the dispersion of DSS, the upper and the lower branches of topological surface state have same band curvatures or carriers velocities. Furthermore, these states retain the saddle-like features of DSS energy dispersion which is more clearly visible in the Fermi band contours shown in Figs. \ref{fig:surface010}(e)-\ref{fig:surface010}(f). The constant energy contours (CECs) are open and disperse along the $k_x$ direction above the Dirac-point or the saddle-point. As we lower the Fermi energy, the electronic states undergo a Lifshitz transition and CECs change its direction and dispersion to lie along the $k_z$ direction. These topological surface state features are unique to ZrPtGe family and provide an excellent materials platform to investigate interaction-induced instabilities on the surface of topological materials.

\begin{figure}[t] 
\includegraphics[width=1.0\textwidth]{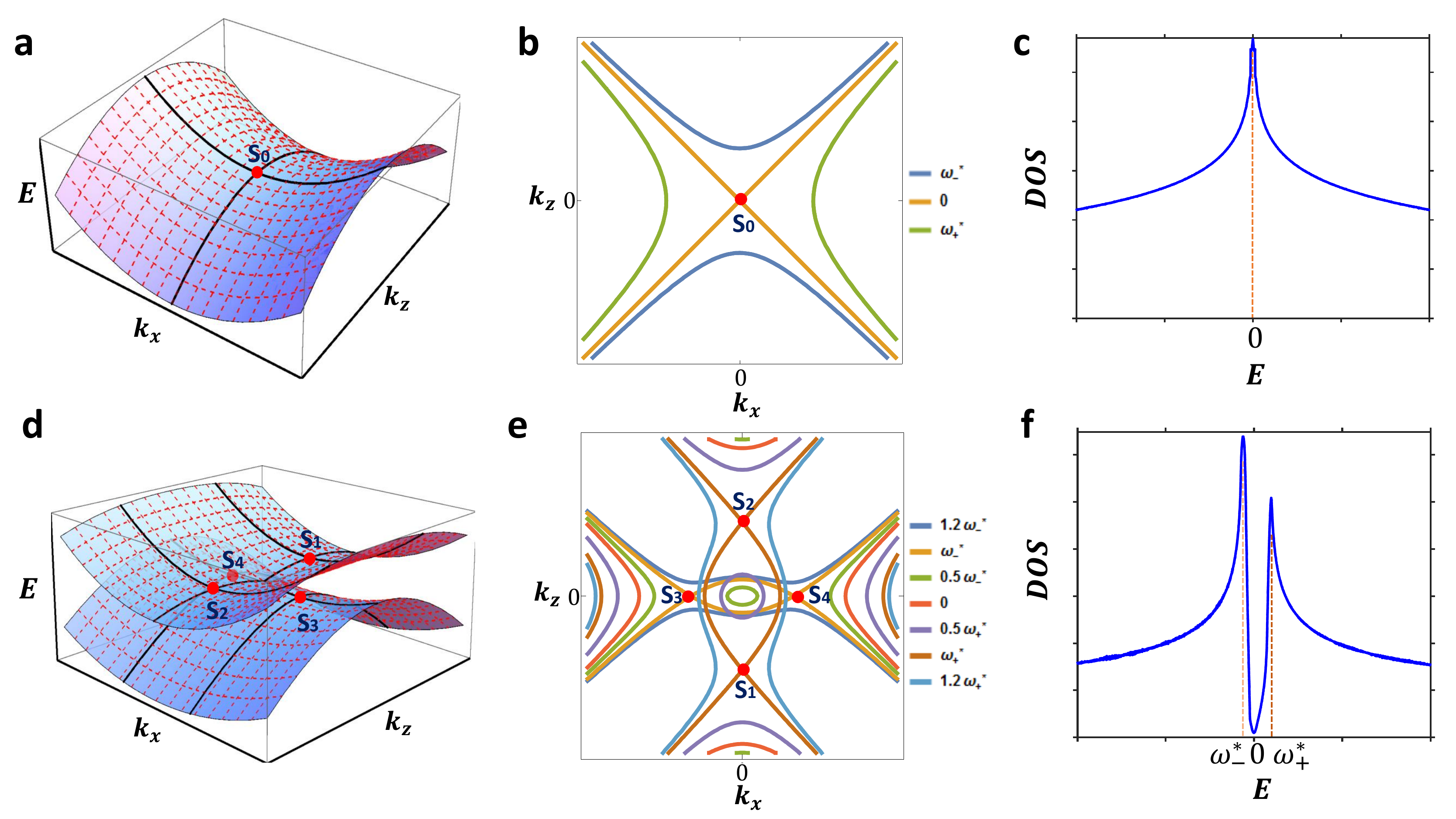} 
\caption{Saddle-like topological surface states. (a) Energy dispersion of the Hamiltonian $ H_{0}(k_{x},k_{z})$ with $\eta=+1$ and (b) associated isoenergy band contours without SOC. A single saddle-point $S_{0}$ of the DSS lies at the $\Gamma$ point. See main text for the meaning of energy band contours $\omega^*_{+}$ and $\omega^*_{-}$. (c) Calculated density of state (DOS) for the surface states using the model Hamiltonian (eq. \ref{H0}). (d)-(e) Same as (a)-(b) but with the inclusion of SOC. Four saddle points away from $\Gamma$ point ($S_1$, $S_2$, $S_3$, and $S_4$) emerge due to the band splitting in presence of SOC. (f) DOS for the surface states with the inclusion of SOC (eq. \ref{Hsoc}). }
\label{fig:saddleSS}
\end{figure}

In order to develop a better understanding of the saddle-like TSSs discussed above, we now present a low-energy effective model Hamiltonian which is obtained using the theory of invariants in a similar way as in the case of Bi$_2$Te$_3$ \cite{Hexawarp_Fu}. On the (010) surface besides $\Theta$, only glide-mirror symmetry $\overline{M}_{z}={\lbrace M_{z}|\frac{1}{2} 0 \frac{1}{2}\rbrace}$ which sends $(k_{x},k_{z}) \rightarrow (k_{x},-k_{z})$ is preserved. Based on our first-principles results and symmetry analysis, a single band $k.p$ model Hamiltonian is enough to describe the DSSs on (010) surface in the absence of SOC. Therefore, the single band effective model Hamiltonian for the surface state can be expressed as

\begin{equation}\label{H0}
H_{0}(k_{x},k_{z}) = \frac{1}{2m^*}(k_{x}^2 - {\eta} k_{z}^2) =\frac{1}{2m^*}[ \frac{1+\eta}{4}(k_{+}^2 + k_{-}^2)+\frac{1-\eta}{2}k_{+}k_{-}],
\end{equation}
where $k_{\pm}=k_{x}\pm ik_{z}$.  Here coefficient $\eta$ determines the form of $E-K$ dispersion relation and depends upon the rotational symmetries present in system  and its material properties. While  $\eta < 0$ gives a paraboloidal energy dispersion, $\eta>0$ ensures a saddle-like energy dispersion. In a system, the n-fold rotational symmetry $C_{n}$ with $n > 2$ normal to the surface forbids $\eta>0$ i.e saddle-like energy dispersion for the surface state whereas $n \le 2$ allows it. This can be verified easily from symmetry constrains of $C_{n}$ on the Hamiltonian $H_{0}(k_{x},k_{z})$ with $\eta > 0$. Since $C_n$ sends $k_{\pm} \rightarrow k_{\pm}e^{\pm i\frac{2\pi}{n}}$ and $(k_{+}^2 + k_{-}^2) \rightarrow (k_{+}^2 e^{i\frac{4\pi}{n}} + k_{-}^2 e^{-i\frac{4\pi}{n}})$, this Hamiltonian remains invariant only for $C_{n}$ with $n \le 2$. It is worth noting that this is only a necessary condition for realizing a saddle-like energy dispersion which depends also on the material properties. As ZrPtGe has a big nodal line energy dispersion in the bulk and its (010) surface lacks $C_{ny}$ with $n>2$, it hosts a symmetry allowed saddle-like state as found in our first-principles results (see Figs. \ref{fig:surface010}). The energy dispersion associated with the model Hamiltonian \ref{H0} is presented in Figs. \ref{fig:saddleSS}(a)-\ref{fig:saddleSS}(c), which further show a saddle-like energy dispersion relation for the DSS with a single saddle-point at the $\Gamma$-point. The density of states (DOS) is logarithmically diverging at the $\Gamma$ point, confirming it as a saddle-point VHS.

In the presence of SOC, the DSS splits into two branches, developing into the spin-polarized surface states of a strong $Z_2$ TI. A two-band $k.p$ Hamiltonian is therefore necessary to describe these states. Considering $\overline{M}_{z} = -e^{-i(k_x-k_z)/2}i\sigma_{z}$ and $\Theta=i\sigma_{y}K$, where $\sigma_{i=x,y,z}$ denote Pauli spin matrices, and $K$ is complex conjugate operator, the effective model Hamiltonian with SOC can be written as

\begin{equation}\label{Hsoc}
H_{SOC}(k_{x},k_{z}) = \frac{1}{2m^*}(k_{x}^2 - {\eta} k_{z}^2)-v_{k}(k_{x}\sigma_{z}-k_{z}\sigma_{x})+v'_{k}k_{z}\sigma_{y},
\end{equation}
where $v_{k}=v_{x}=v_{0}(1+\alpha k^2)$ and ${v_z} = \sqrt{v_{k}^2+{v'_{k}}^2}=v_{z0}(1+\alpha k^2)$ are the Dirac velocities along $x$- and $z$-axis, respectively, with the second order correction. The corresponding energy dispersion relation $E_{\pm}(\textbf{k})$ is
\begin{equation}
E_{\pm}(\textbf{k}) = \frac{1}{2m^*}(k_{x}^2 - {\eta} k_{z}^2) \pm \sqrt{v_{x}^2 k_{x}^2 + v_{z}^2 k_{z}^2}.
\end{equation}

The above energy dispersion demonstrates a new type of symmetry allowed nontrivial topological surface state which is distinct from surface states studied so far\cite{Review_hasan2015,bi2se3_expxia,Hexawarp_Fu}. This new TSS evolves from the saddle-like DSS, having a single Dirac-point at $\Gamma$-point and two pairs of saddle points. The two saddle points are located on $(k_x,k_z)=(\pm m^*v_x,0)$ with energy $\omega^*_{-}= -\frac{m^*v_{x}^2}{2}$, on the lower branch and other two at $(k_x,k_z)=(0, \pm m^*v_z/{\eta})$ with energy $\omega^*_{+}= \frac{m^*v_{z}^2}{2\eta}$ on the higher branch of the TSS (see Figs. \ref{fig:saddleSS}(d)-\ref{fig:saddleSS}(e)). This can be further seen in the DOS shown in Fig. \ref{fig:saddleSS}(f)) where the two saddle-point VHSs are evident at  $\omega^*_{-}$ and  $\omega^*_{+}$. Such unique topological surface states would enable new routes to study novel many body physics to be realized with topological states.

In summary, we propose that the TT'X family of silicides and germanides realize a single topological protected nodal line on the $k_y=0$ plane in the absence of the SOC.  We have shown the existence of DSSs nested inside the nodal line on (010) surface of ZrPtGe as an example with a unique saddle-like energy-momentum dispersion relation. Inclusion of the SOC gaps the nodal line eventually driving the material into the TI phase with $Z_2=(1;000)$. The nontrivial topological surface states evolve from the DSS retaining its saddle-like energy dispersion relation. Our results establish for the first time that TT'X materials family provides an ideal platform to explore properties related with the NLSMs and the saddle-like topological surface states.

This research is supported by the National Research Foundation, Prime Minister's Office, Singapore under its NRF fellowship (NRF Award No. NRF-NRFF2013-03) and benefited from the high performance computing facilities of the Centre for Advanced 2D Materials. The work at Northeastern University was supported by the US Department of Energy (DOE), Office of Science, Basic Energy Sciences grant number DE-FG02-07ER46352 (core research), and benefited from  Northeastern University's Advanced Scientific Computation Center (ASCC), the NERSC supercomputing  center through DOE grant number DE-AC02-05CH11231, and support (applications to layered materials)  from the DOE EFRC: Center for the Computational Design of Functional Layered Materials (CCDM) under DE-SC0012575.

\newpage

\end{document}